# Propagation of Electromagnetic Waves in Extremely Dense Media

**Samina Masood [a] and Iram Saleem [b,*]**

[a] *Department of Physical and Applied Sciences, University of Houston-Clear Lake, Houston, TX, 77058, USA*

[b] *Department of Physics and Texas Center for Superconductivity, University of Houston, Houston, TX, 77204, USA*

[*]Corresponding author email address: masood@uhcl.edu

**Abstract**. We study the propagation of electromagnetic (EM) waves in extremely dense exotic systems with very unique properties. These EM waves develop a longitudinal component due to interactions with the medium. Renormalization scheme of QED is used to understand the propagation of EM waves in both longitudinal and transverse directions. The propagation of EM waves in a quantum statistically treatable medium affects the properties of the medium itself. The electric permittivity and the magnetic permeability of the medium are modified and influence the related behavior of the medium. All the electromagnetic properties of a medium become a function of temperature and chemical potential of the medium. We study in detail the modifications of electric permittivity and magnetic permeability and other related properties of a medium in the superdense stellar objects.

## 1. Introduction

We study the propagation of electromagnetic waves in a hot and dense medium of electrons at extremely high chemical potential. High temperatures with negligible values of chemical potential were available in the very early universe [1-2] whereas extremely large chemical potential with relatively lower temperatures can be found in superdense stars [3-4]. High magnetic fields are expected at high densities. The QED (Quantum Electrodynamics) renormalization scheme in a real-time formalism [5-17] is used to study the vacuum polarization tensor of photons in an extremely dense medium using the longitudinal and transverse components of the vacuum polarization tensor. The basic difference between electromagnetic signals in vacuum and the electromagnetic signals in a medium is due to the longitudinal component of the signal which is developed at the expense of transversality of the electromagnetic waves. The vacuum polarization in the presence of the chemical potential dependent longitudinal component for such a system modifies the electromagnetic properties of the medium itself. We revisit the vacuum polarization tensor of the quanta of energy in a superdense medium and study the electromagnetic properties of the medium including electric permittivity, magnetic permeability, refractive index and the related electromagnetic properties. It is previously found that the neutrino decoupling temperature in the early universe is affected by the temperature as the decoupling process becomes a function of temperature [18-21] at very high temperatures. Therefore the magnetic moment of neutrino gets a thermal contribution with a very tiny mass at the higher loop level due to the coupling of neutrino with the magnetic field. Neutrino magnetic moment also depends on the chemical potential in highly dense systems [22-24]. In general, the electromagnetic properties of an extremely dense system at very high temperature become a function of temperature as well as chemical potential and the corresponding properties of propagating particles are modified in such a medium [25]. In this paper we focus on the study of the electromagnetic properties of a medium due to the perturbative contributions to the vacuum polarization tensor at high temperature and density.

Section 2 is comprised of the detailed properties and discussion of the polarization tensor in the statistical medium. Section 3 is completely devoted to the study of the longitudinal and the transverse components of electromagnetic waves in extremely dense systems. Section 4 is comprised of electromagnetic properties of a superdense system and discuss the behavior of electric permittivity and magnetic permeability at large values of chemical potential. Section 5 is devoted to the quantitative discussion of the analytical results obtained in Sections 3 and 4. Some of the applications of our results to the superdense systems are mentioned in the end also.

## 2. Polarization Tensor

We calculate all the components of vacuum polarization tensor of photon at high temperatures and densities of a medium to study the propagation of light in different directions. Our hypothetical system corresponds to a heat bath of electrons at an equilibrium temperature where density of electrons is much larger than the temperature of the medium. Such systems are available in superdense stellar objects such as supernovae and neutron stars [3-4]. In an extreme situation, density may be much larger than the temperature and we identify such systems as superdense systems hereafter. Such systems have been referred to as classical systems in literature [14-15] because the chemical potential is large enough to dominate the temperature (kinetic energy). Standard classical electrodynamics work perfectly fine in the classical limit so we can use well-known equations of quantum electrodynamics (QED) to study the electric field polarization and the magnetization phenomenon of electrodynamics in a statistical medium. The standard form of vacuum polarization tensor in QED at finite temperature and density (FTD) can be written by replacing the electron propagator in vacuum by the one in the real-time formalism (see, for example Ref [8, 17]).

$$\pi_{\mu\nu}(K,\mu) = ie^2 \int \frac{d^4p}{(4\pi)^4} Tr\{\gamma_\mu(\gamma_\alpha p^\alpha + \gamma_\alpha K^\alpha + m)\gamma_\nu(\gamma_\alpha p^\alpha + m)\}\left[\frac{1}{(p+K)^2 - m^2} + \Gamma_F(p+K,\mu)\right]\left[\frac{1}{p^2 - m^2} + \Gamma_F(p,\mu)\right]$$
(1)

With [8],

$$\Gamma_F(p,\mu) = 2\pi i\delta(p^2 - m^2)[\theta(p_0)n_F(p,\mu) + \theta(-p_0)n_F(p,-\mu)]$$
(2)

Where the 4-momentum of photon K satisfies the relations:

$$K^2 = \omega^2 - k^2, \qquad \omega = K_\alpha u^\alpha.$$
(3)

The vacuum part $\pi_{\mu\nu}^{T=0}(K)$ and the medium contribution $\pi_{\mu\nu}^{\beta}(K,\mu)$ to Eq. (1) can be separated as [5, 14],

$$\pi_{\mu\nu}(K) = \pi_{\mu\nu}^{T=0}(K) + \pi_{\mu\nu}^{\beta}(K,\mu)$$

With,
(4)

$$\pi_{\mu\nu}^{\beta}(K,\mu) = -\frac{2\pi e^2}{2}\int \frac{d^4p}{(4\pi)^4} Tr\{\gamma_\mu(\gamma_\alpha p^\alpha + \gamma_\alpha K^\alpha + m)\gamma_\mu(\gamma_\alpha p^\alpha + m)\}\left[\frac{\delta[(p+K)^2 - m^2]}{p^2 - m^2}\{n_F(p+K,\mu) + n_F(p+K,-\mu)\} + \frac{\delta[p^2 - m^2]}{(p+K)^2 - m^2}\{n_F(p,\mu) + n_F(p,-\mu)\}\right].$$
(5)

The polarization tensor $\pi_{\mu\nu}^{\beta}(K,\mu)$ can generally be written in terms of the longitudinal and transverse components $\pi_L(k,\omega)$ and $\pi_T(k,\omega)$, respectively, such that

$$\pi_{\mu\nu}(K,\mu) = P_{\mu\nu}\pi_T(K,\mu) + Q_{\mu\nu}\pi_L(K,\mu) \tag{6}$$

Where

$$P_{\mu\nu} = \tilde{g}_{\mu\nu} + \frac{\tilde{K}_\mu \tilde{K}_\nu}{k^2} \tag{7a}$$

and

$$Q_{\mu\nu} = -\frac{1}{K^2 k^2}(k^2 u_\mu + \omega \tilde{K}_\mu)(k^2 u_\nu + \omega \tilde{K}_\nu) \tag{7b}$$

Whereas,

$$\tilde{g}_{\mu\nu} = g_{\mu\nu} - u_\mu u_\nu \tag{7c}$$

and

$$\tilde{K}_\mu = K_\mu - \omega u_\mu \tag{7d}$$

Such that they satisfy the conditions:

$$P_\nu^\mu P_\alpha^\mu = P_\alpha^\mu \qquad Q_\alpha^\mu = Q_\nu^\mu Q_\alpha^\nu \qquad K_\mu P_\nu^\mu = 0 \qquad K_\mu Q_\nu^\mu = 0 \tag{8}$$

From the above relations, we can write

$$P_{\mu\nu} = \begin{pmatrix} 0 & 0 & 0 & 0 \\ 0 & -1 - \frac{k_1^2}{k^2} & -\frac{k_1 k_2}{k^2} & -\frac{k_1 k_3}{k^2} \\ 0 & -\frac{k_1 k_2}{k^2} & -1 - \frac{k_2^2}{k^2} & -\frac{k_2 k_3}{k^2} \\ 0 & -\frac{k_1 k_3}{k^2} & -\frac{k_2 k_3}{k^2} & -1 - \frac{k_3^2}{k^2} \end{pmatrix} \tag{9a}$$

and

$$Q_{\mu\nu} = \begin{pmatrix} -\frac{k^2}{K^2} & -\frac{i\omega k_1}{K^2} & -\frac{i\omega k_2}{K^2} & -\frac{i\omega k_3}{K^2} \\ -\frac{i\omega k_1}{K^2} & \frac{\omega^2 k_1^2}{k^2 K^2} & \frac{\omega^2 k_1 k_2}{k^2 K^2} & \frac{\omega^2 k_1 k_3}{k^2 K^2} \\ -\frac{i\omega k_2}{K^2} & \frac{\omega^2 k_1 k_2}{k^2 K^2} & \frac{\omega^2 k_2^2}{k^2 K^2} & \frac{\omega^2 k_2 k_3}{k^2 K^2} \\ -\frac{i\omega k_3}{K^2} & \frac{\omega^2 k_1 k_3}{k^2 K^2} & \frac{\omega^2 k_2 k_3}{k^2 K^2} & \frac{\omega^2 k_3^2}{k^2 K^2} \end{pmatrix} \tag{9b}$$

Where $P_{\mu\nu}$ and $Q_{\mu\nu}$ are the polarization tensors associated with the longitudinal and transverse components respectively. The complete polarization tensor is then expressed in the form of longitudinal and transverse components as

$$\pi_{\mu\nu} = \begin{pmatrix} -\frac{k^2}{K^2}\pi_L & -\frac{i\omega k_1}{K^2}\pi_L & -\frac{i\omega k_2}{K^2}\pi_L & -\frac{i\omega k_3}{K^2}\pi_L \\ -\frac{i\omega k_1}{K^2}\pi_L & \left(-1-\frac{k_1^2}{k^2}\right)\pi_T + \left(\frac{\omega^2 k_1^2}{k^2 K^2}\right)\pi_L & \left(-\frac{k_1 k_2}{k^2}\right)\pi_T + \left(\frac{\omega^2 k_1 k_2}{k^2 K^2}\right)\pi_L & \left(-\frac{k_1 k_3}{k^2}\right)\pi_T + \left(\frac{\omega^2 k_1 k_3}{k^2 K^2}\right)\pi_L \\ -\frac{i\omega k_2}{K^2}\pi_L & \left(-\frac{k_1 k_2}{k^2}\right)\pi_T + \left(\frac{\omega^2 k_1 k_2}{k^2 K^2}\right)\pi_L & \left(-1-\frac{k_2^2}{k^2}\right)\pi_T + \left(\frac{\omega^2 k_2^2}{k^2 K^2}\right)\pi_L & \left(-\frac{k_2 k_3}{k^2}\right)\pi_T + \left(\frac{\omega^2 k_2 k_3}{k^2 K^2}\right)\pi_L \\ -\frac{i\omega k_3}{K^2}\pi_L & \left(-\frac{k_1 k_3}{k^2}\right)\pi_T + \left(\frac{\omega^2 k_1 k_3}{k^2 K^2}\right)\pi_L & \left(-\frac{k_2 k_3}{k^2}\right)\pi_T + \left(\frac{\omega^2 k_2 k_3}{k^2 K^2}\right)\pi_L & \left(-1-\frac{k_3^2}{k^2}\right)\pi_T + \left(\frac{\omega^2 k_3^2}{k^2 K^2}\right)\pi_L \end{pmatrix} \quad (10)$$

Eq. (10) shows that $\pi_{\mu 0}$ and $\pi_{0\nu}$ are all proportional to $\pi_L$ and have imaginary negative values except $\pi_{00} = -\frac{k^2}{K^2}\pi_L$. Moreover there is a complete symmetry between off-diagonal components at a given temperature. Reflection symmetry is still maintained but spherical symmetry may be compromised.

## 3. Longitudinal and Transverse Components

The longitudinal and transverse components of the vacuum polarization tensors have already been calculated in different limits of temperature and chemical potential.

(i) For $T \gg m \gg \mu$ we can write the longitudinal and transverse component for the same particle-antiparticle concentration and same chemical potential with opposite sign in the Fermi-Dirac distribution functions [4, 5] as

$$\pi_L \cong \frac{4e^2}{\pi^2}\left(1 - \frac{\omega^2}{k^2}\right)\left[\left(1 - \frac{\omega}{2k}\ln\frac{\omega+k}{\omega-k}\right)\left(\frac{ma(m\beta,\mu)}{\beta} - \frac{c(m\beta,\mu)}{\beta^2}\right) \right. \\ \left. + \frac{1}{4}\left(2m^2 - \omega^2 + \frac{11k^2 + 37\omega^2}{72}\right)b(m\beta,\mu)\right] \quad (11a)$$

$$\pi_T \cong \frac{2e^2}{\pi^2}\left[\left\{\frac{\omega^2}{k^2} + \left(1 - \frac{\omega^2}{k^2}\right)\ln\frac{\omega+k}{\omega-k}\right\}\left(\frac{ma(m\beta,\mu)}{\beta} - \frac{c(m\beta,\mu)}{\beta^2}\right) \right. \\ \left. + \frac{1}{8}\left(2m^2 + \omega^2 + \frac{107\omega^2 - 131k^2}{72}\right)b(m\beta,\mu)\right] \quad (11b)$$

In a CP- symmetric background, with equal chemical potentials for particles and antiparticles, we can write Masood's *abc* functions in Eq. (11) for $T \gg \mu$ as

$$a(m\beta, \pm\mu) = \ln(1 + e^{-\beta(m\pm\mu)})$$

$$b(m\beta, \pm\mu) = \sum_{n=1}^{\infty} (-1)^n e^{\mp\beta\mu} Ei(-nm\beta) \quad (12)$$

$$c(m\beta, \pm\mu) = \sum_{n=1}^{\infty}(-1)^n \frac{e^{-n\beta(m\pm\mu)}}{n^2}$$

Masood's functions in Eq. (12) can be averaged on particle-antiparticle contributions in a CP symmetric background and can also be evaluated individually for particles or antiparticles, as needed.

(ii) In the limit $T \ll m \ll \mu$

$$\pi_L(K,T,\mu) \cong \frac{e^2}{2\pi^2 k}\left(1 - \frac{\omega^2}{k^2}\right)\left[\left(k - 2\omega \ln\frac{\omega-k}{\omega+k}\right)J_1 - 2A_1 J_2 - \frac{1}{2}A_2 J_3\right] \quad (13a)$$

$$\pi_T(K,T,\mu) \cong \frac{e^2}{2\pi^2 k}\left[\left\{\frac{\omega^2}{k} + \omega\left(1 - \frac{\omega^2}{k^2}\right)\ln\frac{\omega+k}{\omega-k}\right\}J_1 + \left\{\frac{kK^2 + km^2}{2} + A_1\left(1 - \frac{\omega^2}{k^2}\right)\right\}J_2 \right.$$
$$\left. + \left\{\frac{k^3}{12} + \frac{K^2 k}{24m^2}(3\omega^2 + k^2)\right\} + \left(1 - \frac{\omega^2}{k^2}\right)\frac{A_2}{4m^2} - \frac{m^2(5\omega^2 - 3k^2)}{8k}J_0\right] \quad (13b)$$

Where

$$A_1 = \frac{k^3}{12} + \frac{\omega^2 k}{2} + \frac{m^2 k}{K^2}(\omega^2 + k^2) + \frac{kK^2}{2} - \omega^2 k\left[\frac{2m^2}{k^2} - 1\right] \quad (14a)$$

$$A_2 = \frac{1}{4}\left[\frac{m^2}{K^4}\{4m^2 k^3(4\omega^2 + k^2) - K^2(3k\omega^4 + 12\omega^2 k^3 + k^5)\} + k\left[\omega^4 + 20\omega^2 k^2 + 2k^4 + \frac{m^4}{2}\right] + \frac{K^2 k}{6}(3\omega^2 + k^2)\right] \quad (14b)$$

And

$$J_1 = \frac{1}{2}\left[\frac{\mu^2}{2}\left[1 - \frac{m^2}{\mu^2}\right] + \frac{1}{\beta}\{a(m\beta,\mu) - a'(m\beta,\mu)\} - \frac{1}{\beta^2}\{c(m\beta,\mu) + c'(m\beta,\mu)\}\right] \quad (15a)$$

$$J_2 = \frac{1}{2}\left[\ln\frac{\mu}{m} + b(m\beta,\mu) - b'(m\beta,\mu)\right] \quad (15b)$$

$$J_3 = \frac{1}{2}\left[\frac{1}{2m^2}\left[1 - \frac{m^2}{\mu^2}\right] - \frac{1}{4\mu^2} + \frac{1}{m^2}\{n_f(\mu+m) + n_f(\mu-m)\}\right.$$
$$\left. + \frac{\beta}{m}\left\{\frac{e^{-\beta(\mu+m)}}{[1 + e^{\beta(\mu+m)}]^2} + \frac{e^{-\beta(\mu-m)}}{[1 + e^{-\beta(\mu-m)}]^2}\right\} + d(m\beta,\mu) + d'(m\beta,\mu)\right] \quad (15c)$$

Nonzero values of the longitudinal component in the above equations show that the quanta of electromagnetic waves acquire dynamically generated mass in a medium and due to this effective mass it generates a longitudinal component which affects the electromagnetic properties of the medium. The difference between the values of $\pi_L$ and $\pi_T$ in Eqs (11) and Eqs (13) correspond to the dynamically generated mass of photon.

## 4. Electromagnetic Properties

Modification of electromagnetic properties of a medium are indebted to the longitudinal component of the polarization tensor and is identified by the electromagnetic properties of the medium. Some of these electromagnetic properties are derived from the vacuum polarization tensor. We compute the electric permittivity, $\varepsilon(k)$ magnetic permeability $\mu(k)$, and the refractive index $r_i$ in a superdense medium as:

$$\varepsilon(k) = 1 - \frac{\pi_L(K)}{k^2} \qquad \frac{1}{\mu(k)} = \frac{k^2\pi_T(K) - \omega^2\pi_L(K)}{k^2K^2} \tag{16a}$$

$$v_{prop} = \sqrt{\frac{1}{\varepsilon(K)\mu(K)}} \qquad r_i = \frac{c}{v} = \sqrt{\frac{\varepsilon(K)\mu(K)}{\varepsilon_0(K)\mu_0(K)}} \tag{16b}$$

Substituting the values of $\pi_L(K)$ and $\pi_T(K)$ in Eq. (16), we can determine the corresponding values of $\varepsilon(k)$ and $\mu(k)$ for different regions of temperature and chemical potential. We evaluate these results for extreme conditions only. It has already been worked out [13, 17-18] at $T \gg \mu$ and can be evaluated for some favorable conditions as:

(i) $\omega \gg k$

$$\pi_L = -\frac{\omega^2 e^2 T^2}{3k^2} \qquad \pi_T = \frac{\omega^2 e^2 T^2}{6k^2} \tag{17}$$

$$\varepsilon(k) = 1 + \frac{\omega^2 e^2 T^2}{3k^4} \qquad \frac{1}{\mu(k)} \cong \frac{\omega^4 e^2 T^2}{3k^4 K^2} \tag{18}$$

(ii) $\omega \ll k$

$$\pi_L = \frac{e^2 T^2}{3} \qquad \pi_T = \frac{e^2 T^2}{6} \tag{19}$$

$$\varepsilon(k) = 1 - \frac{e^2 T^2}{6k^2} \qquad \frac{1}{\mu(k)} \cong \frac{e^2 T^2}{6K^2} \tag{20}$$

Above equations show that the contribution at high temperatures (of the order T²) is due to $c(m\beta, \mu)$ which reduces to a factor ($-\pi^2/12$) at extremely high temperatures. The limits on the coefficients $A_1$ and $A_2$ can be imposed by the comparison of the parameters $\omega$ and $k$. A massless photon is transverse in nature, i.e $\omega^2 = k^2$. However, when the photon acquires dynamically generated mass in a hot and dense medium, we can evaluate the coefficients $A_1$ and $A_2$ corresponding to suitable limits of $\omega$ and $k$.

(i) Where $\omega \gg k$, $K^2 = \omega^2$ and $\frac{k}{\omega} \ll 1$ ($k^2$ is ignored as compared to $\omega^2$)

$$A_1 = \frac{-2m^2\omega^2}{k} \quad \text{and} \quad A_2 = \frac{3\omega^4 k}{8} \tag{21a}$$

(ii) Where $\omega \ll k$, $K^2 = -k^2$ and $\frac{\omega}{k} \ll 1$

$$A_1 = \frac{-k^3}{3} \quad \text{and} \quad A_2 = \frac{11k^5}{24} \tag{21b}$$

For $\mu \gg T$, it can be easily seen that $a' = \mu ln2$ and is constant for constant $\mu$ whereas Masood's other functions have negligibly small contributions, such that:

$$J_1 = \frac{1}{2}\left[\frac{\mu^2}{2} - \frac{m^2}{2}\right]$$

$$J_2 = \frac{1}{2}[ln\frac{\mu}{m}] \tag{22}$$

$$J_3 = \frac{1}{4m^2}$$

Substitution of Eq. (15) and Eq. (16) in Eq. (13) gives the electric permittivity and magnetic permeability as:

$$\varepsilon_E(k,\mu,T) = 1 - \frac{2\alpha}{\pi K^2 k}\left[\left[k - 2\omega ln\frac{\omega-k}{\omega+k}\right]J_1 - 2A_1 J_2 + \frac{1}{2}A_2 J_3\right]\left[1 - \frac{\omega^2}{k^2}\right] \tag{23a}$$

$$\frac{1}{\mu_B(k,\mu,T)} \cong 1 + \frac{2\alpha}{\pi K^2 k^3}\left[\begin{array}{c}\left[\frac{k^2}{2}+\omega^2\right]\left[1-\frac{\omega^2}{k^2}\right]\left\{2\omega ln\frac{\omega-k}{\omega+k}J_1 - 2A_1 J_2 - \frac{1}{2}A_2 J_3\right\} \\ -\frac{k^3 K^2 + k^4 m}{4}J_2 - \frac{m^2}{2}\left\{\frac{k^5}{12} + \frac{(3\omega^2+k^2)k^3 K^2}{24m^2} + \frac{m^2 k^2(5\omega^2 - 3k^2)}{8k}\right\}J_3\end{array}\right] \tag{23b}$$

Therefore, just to understand the difference of behavior, we consider extremely dense situations to quantitatively analyze our results. We basically consider two extreme conditions:

1. For $\mu \gg T$ and $\omega \gg m \gg k$, we obtain from the above equations

$$\varepsilon_E(k,\mu) = 1 + \frac{2\alpha}{\pi^2}\left[\frac{\mu^2}{4k^2} + \frac{2m^2\omega^2}{k^4}ln\frac{\mu}{m}\right] \tag{24a}$$

$$\frac{1}{\mu_B(k,\mu)} = 1 + \frac{2\alpha}{\pi}\frac{\omega^6}{k^6}\left[\frac{3}{16}\frac{k^2}{4m^2} - \frac{2m^2}{\omega^2}ln\frac{\mu}{m}\right] \tag{24b}$$

Eq. (24a) shows that the dominant radiative contribution to electric permittivity comes from the first term in parenthesis which is a quadratic term; the ratio between the chemical potential and the momentum $k$. Second term in the parenthesis is $ln\frac{\mu}{m}$ a slowly varying function of chemical potential and does not contribute much. Due to this term, the magnetic permeability (Eqn. (24)) correction does not strongly depend on the chemical potential. Permeability gets a significant contribution from the energy due to its strong dependence coming from the term $\frac{\omega^6}{m^2 k^4}$.

2. For $\mu \gg T$ and $\omega \ll m \ll k$

$$\varepsilon_E(k,\mu) = 1 + \frac{\alpha}{2\pi^2}\left[\frac{\mu^2}{k^2} + \frac{11}{48}\frac{k^2}{m^2}\right] \tag{25a}$$

$$\frac{1}{\mu_B(k,\mu,T)} = 1 + \frac{\alpha}{2\pi}\left[\frac{3}{32}\frac{k^2}{m^2}\right] \tag{25b}$$

Chemical potential dependence of electric permittivity in case (2) is similar to case (1) and is a quadratic function of $\mu$. However, in this case, the $k/m$ dependence is obvious. Both of these cases are discussed in detail in the next section.

### 4. Results and Discussion

For electromagnetic waves, k corresponds to the momentum of a propagating wave and ω corresponds to its energy. For extremely large values of μ, two possible conditions are ω > k or k > ω. We consider both of these cases one by one and see the dependence of electric permittivity and magnetic permeability on k and ω for different values of chemical potential. For superdense astrophysical systems like neutron stars, the chemical potential could be as high as a few hundred MeV [2] or so. Therefore we have included a typical case of μ=500 m which corresponds to approximately 250 MeV energy.

We mainly analyze Eq. (25) and Eq. (26) in this section and plot them for different values of μ, ω and k. Figure (1) gives a plot of electric permittivity as a function of ω, the energy of the photon. We plot it for different values of μ. It is obvious from the plot that the μ dependence changes the value of permittivity for lower energies, typically below ω=3m, which corresponds to E < 1.5 MeV. When E >1.5 MeV the electric permittivity vanishes indicating the trapping of light or opacity of the medium. On the other hand a small region of negative permittivity corresponds to the behavior of metamaterials where the material may be invisible [26]. However if the value of k, the photon momentum is maintained to be constant with the increase in $\omega$ (energy of the photon), this behavior may not be seen.

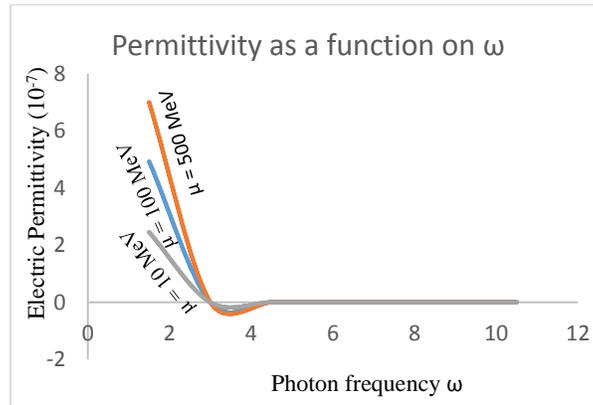

*Figure 1: The graph of electric permittivity $\varepsilon_E$ vs angular frequency $\omega$ for chemical potential values $\mu = 10, 100$ and $500$ MeV in the case of large angular frequency.*

The magnetic permeability, however, has a very different behavior. Figure 2a shows that the magnetic permeability can blow up around 3 MeV energy for large values of k > ω. Whereas, for larger values of the chemical potential it seems to give negative values at 1.75 MeV, which is even below the neutrino decoupling. Magnetic permeability indicates the ability of a material to magnetize. So the increase in chemical potential with negative permeability will reduce the magnetization of material. However, further increase in chemical potential will fully magnetize the material and further magnetization will not be possible.

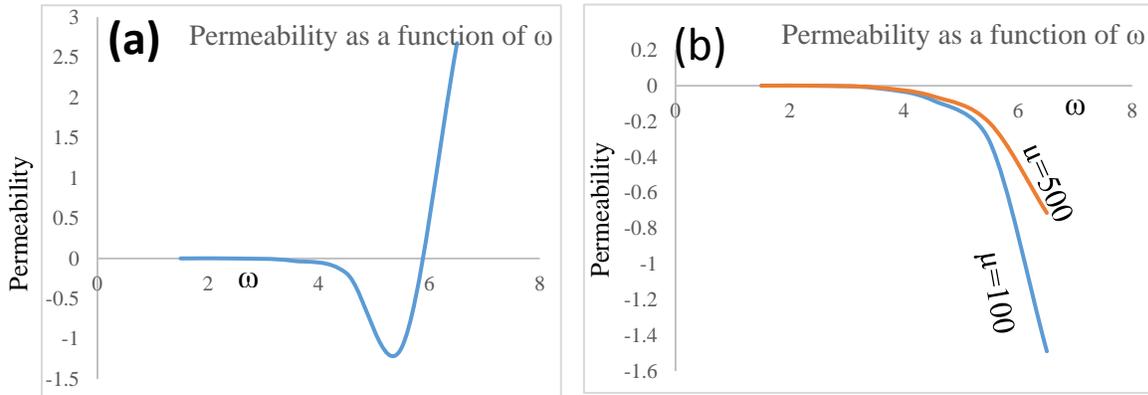

*Figure 2a: The graph of magnetic permeability $\mu_B$ vs angular frequency $\omega$ for chemical potential $\mu = 10$ in the case of extremely large angular frequency. Figure 2b: shows the same graph for $\mu = 100$ and $500$.*

Therefore, for smaller values of chemical potential, magnetic permeability is 1 whereas the permittivity is larger than 1, showing that the medium is conducting. For large values of μ, background corrections to permittivity goes to zero but permeability becomes negative indicating that the magnetic field associated with the high density of electrons can generate some unusual characteristics of the material. The negative permeability under special conditions is an indication of left handedness of some nanofabricated materials, as has already been noticed [24]. The behavior of electromagnetic properties changes with the photon momentum in a relatively different way. Figure (3a) shows that the permittivity of the medium for large k is slowly varying function of photon momentum, however, for large values of the chemical potential, permittivity is significantly increased for lower values of k and attains a constant value around 3.5 MeV.

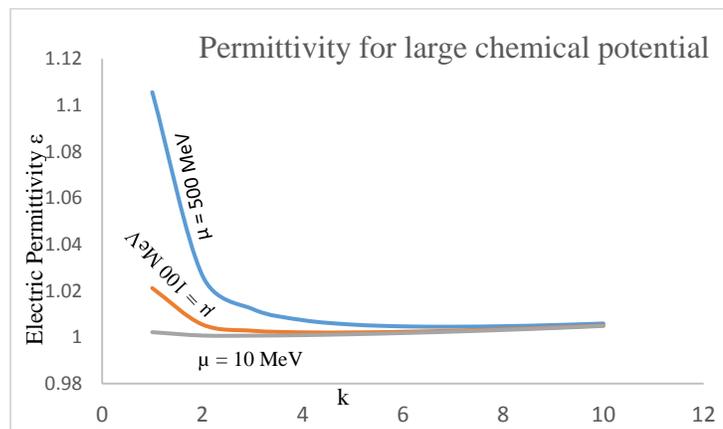

*Figure 3: Shows the graph of electric permittivity $\epsilon_E$ vs the propagation vector $k$ for chemical potential values $\mu = 10, 100$ and $500 MeV$ in the case of extremely large propagation vector.*

Magnetic permeability is not a function of the chemical potential anymore. The change in behavior of permability as a function of k is plotted in Figure 4b.

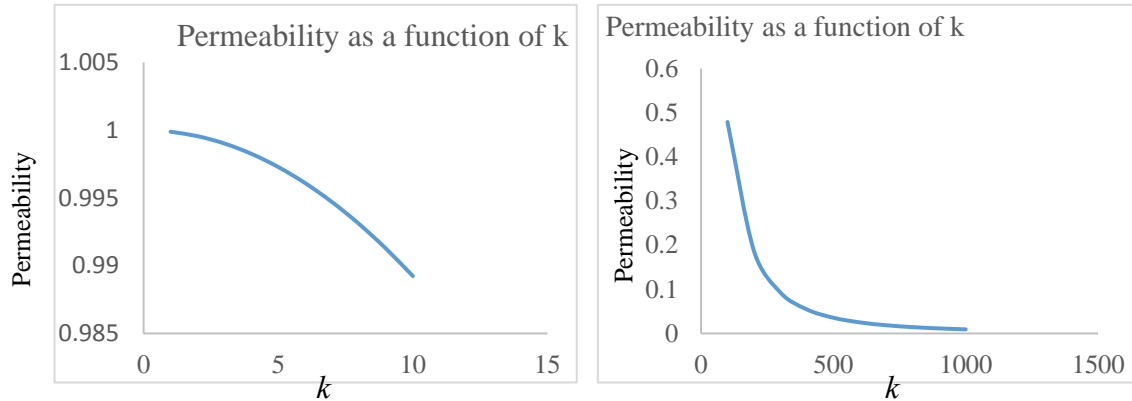

*Figure 4a: The graph of magnetic permeability $\mu_B$ vs propagation vector k in the case of extremely large propagation vector k for the range k =1 to 10. The magnetic permeability in this case is independent of the chemical potential $\mu$. Figure 4b: shows the same graph for the range k =100 to 1000.*

The overall behavior of magnetic permeability for a large range of momentum is plotted as a function of k in Figure 4. This figure shows that the permeability generally varies differently in different regions of k. Fig. (4a) shows that the permeability decreases with k for small values of k. For 150 MeV < k < 450 MeV (Fig.(4b)), it decreases exponentially. For larger than 500 k (around 250 MeV) the magnetic permeability seem to vanish. This type of situation may be availble in the cores of superdense stars such as neutron star core (27-29). For small ranges (such as k=1-10), the local behavior may be a little different (Figure (4b)).

**References and Footnotes**